\documentstyle[aps]{revtex}

\input{tcilatex}

\begin{document}
\title{Deterministic secure communication without using entanglement}
\author{CAI Qing-Yu ( )\thanks{%
Corresponding author. Tel: +862787199109(O); fax: +862787198200 E-mail:
qycai@wipm.ac.cn} and LI Bai-Wen ( )}
\address{Wuhan Institute of Physics and Mathematics, The Chinese Academy of\\
Sciences, Wuhan, 430071, People's Republic of China}
\maketitle

\begin{abstract}
We show a deterministic secure direct communication protocol using single
qubit in mixed state. The security of this protocol is based on the security
proof of BB84 protocol. And it can be realized with today's technology.

\begin{itemize}
\begin{description}
\item  PACS numbers: 03.67.Dd, 03.67.Hk, 03.67.-a
\end{description}
\end{itemize}
\end{abstract}

Quantum key distribution (QKD) is a protocol which is $provably$ secure, by
which private key bit can be created between two parties over a public
channel. The key bits can then be used to implement a classical private key
cryptosystem, to enable the parties to communicate securely. The basic idea
behind QKD is that Eve cannot gain any information from the qubits
transmitted from Alice to Bob without disturbing their states. First, the
no-cloning theorem forbids Eve to perfectly clone Alice's qubit. Secondly,
in any attempt to distinguish between two non-orthogonal quantum states,
information gain is only possible at the expense of introducing disturbance
to the signal [1].

Based on the postulate of quantum measurement [2] and no-cloning theorem
[3], different QKD protocols are presented [4-7]. However, these types of
cryptographic schemes are usually nondeterministic. In Ref.[8], K.
Bostr\"{o}m and T. Felbinger presented a protocol, which allows for
deterministic communication using entanglement. The basic idea of the
ping-pong protocol is that one can encode the information locally on an EPR
pair, but it has a nonlocal effect. In this paper, we show a secure
communication protocol which is a deterministic secure direct communication
protocol using single qubit in mixed state.

This protocol is based on the property that non-orthogonal quantum states
cannot be reliably distinguished [2]. One cannot simultaneously measure the
polarization of a photon in the vertical-horizontal basis and simultaneously
in the diagonal basis. It is well known one can prepare a photon in states $%
\{|0>,|1>\}$ or $\{|\varphi _{0}>,|\varphi _{1}>\}$in its polarization
degree of freedom, where 
\begin{eqnarray*}
|\varphi _{0} &>&=\frac{1}{\sqrt{2}}(|0>+|1>), \\
|\varphi _{1} &>&=\frac{1}{\sqrt{2}}(|0>-|1>).
\end{eqnarray*}
Denoting that $i\sigma _{y}=|0><1|-|1><0|$, it can be obtained: 
\[
i\sigma _{y}|0>=-|1>,\text{ }i\sigma _{y}|1>=0, 
\]
and 
\[
i\sigma _{y}|\varphi _{0}>=|\varphi _{1}>\text{, }i\sigma _{y}|\varphi
_{1}>=-|\varphi _{0}>. 
\]
Suppose Alice want to obtain some information from Bob. First Alice selects
state $|0>$ or $|\varphi _{0}>$ randomly with the probability $\frac{1}{2}$
every time. For an external person without Alice's a prior knowledge, this
qubit appears to be in a mixed state $\rho _{0}:$%
\[
\rho _{0}=\frac{1}{2}|0><0|+\frac{1}{2}|\varphi _{0}><\varphi _{0}|. 
\]
Then Alice sends this qubit to Bob. Bob decides either to perform the
operation $i\sigma _{y}$ on the travel qubit to encode the information `1'
or do nothing, i.e., to perform the operation $I=|0><0|+|1><1|$ to encode
the information `0'. Then Bob sends the travel qubit back to Alice. Alice
performs a measurement on this back qubit to gain the information Bob
encoded. After Alice's decoding measurement, she tells Bob she has received
the back qubit through the public channel by one bit (This can be called as $%
Alice^{\prime }s$ $receipt$. ). In this protocol, there are two
communication modes, `$message$ $mode$' and `$control$ $mode$'. By default,
Bob and Alice are in message mode and the communication is described as
above. With probability $c$, Bob switches the message mode to control mode. $%
In$ $Control$ $Mode$. Instead of his encoding operation, Bob replaces the
qubit he receives from Alice with a qubit that he completely randomly
prepares in the state $|0>$ , $|1>$, $|\varphi _{0}>$ or $|\varphi _{1}>$
and sends this qubit to Alice. Alice performs her decoding measurement and
in 50\% of the case she uses the basis as Bob used. After Alice announces
her receipt of the qubit, Bob announces that this has been a control run and
he tells Alice which state he prepared. If Alice used the same basis as Bob
and if she found a state different from Bob prepared then Eve is detected
and the communication stops. This protocol can be described explicitly like
this:

\begin{quote}
(1). Alice prepares one qubit in state $|0>$ or $|\varphi _{0}>$ randomly
with record.

(2). Alice sends this qubit to Bob.

(3). Bob receives the travel qubit. He decides to the message mode (4m) or
the control mode (4c) by chance.

(4c). $Control$ $mode$. Bob replaces the qubit he receives from Alice with a
qubit that he randomly prepares in the state $|0>$ , $|1>$, $|\varphi _{0}>$
or $|\varphi _{1}>$ and sends this qubit to Alice. Alice performs her
decoding measurement. After Alice announces her receipt of the qubit, Bob
announces that it is a control run this time and he tells Alice which state
he prepared. If Alice used the same basis as Bob and if she found a state
different from Bob prepared then Eve is detected and the communication
stops. Else, Alice sends next qubit to Bob.

(4m). $Message$ $mode$. Bob performs an operation on the travel qubit to
encode information. He encodes the bit `0' using by the operation $I$ and
the bit `1' by the operation $i\sigma _{y}$. Then Bob sends this travel
qubit back to Alice. Alice measures the qubit to gain the message Bob
encoded and sends her $receipt$ to Bob through public channel.

(5). When all of Bob's information is transmitted, this communication is
successfully terminated.
\end{quote}

$Security$ $proof$. The basic idea behind QKD is the fundamental
proposition: Eve can not gain any information from the qubits transmitted
from Alice to Bob with out disturbing their state [1]. Consider that $%
|\varphi >$ and $|\phi >$ are non-orthogonal quantum states Eve is trying to
obtain information about. Without loss of generality that the process she
uses to obtain information is to unitarily interact the state with an
ancilla prepared in a standard state $|e_{0}>$. Assuming that this process
does not disturb the states, then one obtains: 
\begin{eqnarray*}
U|\varphi &>&|e_{0}>=|\varphi >|e_{1}>, \\
U|\phi &>&|e_{0}>=|\phi >|e_{2}>.
\end{eqnarray*}
To acquire information about the different state, Eve would like $|e_{1}>$
and $|e_{2}>$ different. Since the inner products are preserved under
unitary transformations, it must be that 
\[
<\varphi |\phi ><e_{0}|e_{0}>=<\varphi |\phi ><e_{2}|e_{3}>. 
\]
Since $|\varphi >$ and $|\phi >$ are non-orthogonal, then it has 
\[
<e_{2}|e_{3}>=<e_{0}|e_{0}>=1, 
\]
which implies that $|e_{2}>$ and $|e_{3}>$ must be identical. That
distinguishing two non-orthogonal states would at least disturb one of them.

To gain information, Eve has to know which operation Bob performed. First,
she can attack the travel qubit in the line $A\rightarrow B$. And perform a
measurement to acquire Bob's information in line $B\rightarrow A$. Or she
can take another strategy that she only performs a measurement after Bob's
operation to gain Bob's information. No matter what strategy Eve uses, she
has to attack the qubit in the line $B\rightarrow A$. We can see that our
protocol in control mode is as the same as the BB84 protocol's detection of
Eve's eavesdropping. Many works have been accomplished of the security proof
of the BB84 protocol [7, 9]. Eve's any attempt to eavesdrop the information
will give a detection probability $d>0$. Taking into account the probability 
$c$ of a control run, the effective transmission rate is $r=1-c$. The
probability of Eve's eavesdropping one message transfer without being
detected is [8] 
\[
s(c,d)=\frac{1-c}{1-c(1-d)}, 
\]
where $d(I_{0})$ is the detection probability in the control mode. After $n$
protocol run, the probability to successfully eavesdrop $I=nI_{0}(d)$, the
probability to successfully eavesdrop becomes 
\[
s(n,c,d)=(\frac{1-c}{1-c(1-d)})^{I/I_{0}}. 
\]
For $c>0$, $d>0$, this value decreases exponentially. In the limit $%
n\rightarrow \infty $, we have $s\rightarrow 0$. So this protocol is
asymptotically secure. In principle, the security can arbitrarily be
improved by increasing the control parameter $c$ at cost of decreasing the
transmission rate. In practice, we can use a small $c$, which will improves
the efficiency of the communication. To realize a perfectly secure
communication, we must abandon the direct transfer in favor of a key
transfer [8]. Instead of transmitting the message directly to Alice, Bob
will take a random sequence of $N$ bits from a secret random number
generator. After a successful transmission, the random sequence is used as a
shared secret key between Bob and Alice. Bob and Alice can choose classical
privacy amplification protocols, which make it very hard to decode parts of
the message with only some of the key bits given. So Eve has virtually no
advantage in eavesdropping only a few bits. When Eve is detected, the
transfer stops. Then Eve has nothing but a sequence of nonsense random bits
[13].

In contrast to quantum key distribution protocol BB84 [4], our protocol
provides a deterministic transmission of bits. It is possible to communicate
the message directly from Bob to Alice. Essentially, this protocol is a
special case of BB84 protocol. The essence of this protocol is that the
communicators can freely select the message mode and the control mode. In
BB84 protocol, when Alice and Bob want to transform $n$ bit message, it need
about $4n$ qubit. On the other hand, comparing with the `ping-pong'
protocol, we use a single qubit to realize the deterministic secure direct
communication instead of using entanglement. Also, there maybe a
denial-of-service(DoS) attack in the line $A\rightarrow B$ [14]. But any
method of message authentification can protect the protocol against
man-in-the-middle attacks with a reliable public channel.

In order to be practical and secure, a quantum key distribution scheme must
be based on existing---or nearly existing---technology [15]. Experimental
quantum key distribution was demonstrated for first time by Bennett, et al
[16]. Since then, single photon source have been studied in recent years and
a great variety of approaches has been proposed and implemented [17-22].
Today, several groups have shown that quantum key distribution is possible,
even outside the laboratory. In principle, any two-level quantum system
could be used to implement quantum cryptography (QC). In practice, all
implementations have relied on photons. The reason is that their decoherence
can be controlled and moderated. The technological challenges of the QC are
the questions of how to produce single photons, how to transmit them, how to
detect single photons, and how to exploit the intrinsic randomness of
quantum processes to build random generators [23]. Considered the
experimental feasibility, our protocol needs a single photon source and some
linear optical elements and a single-photon detector. Recently, the full
implementation of a quantum cryptography protocol using a stream of a single
photon pulses generated by a stable and efficient source operating at room
temperature was reported [24]. The single pulses are emitted on demand and
the secure bit rate is 7700bits/s. And quantum logic operations using linear
optical elements can be realized with today's technology [25]. The
implementation of the single-photon detection technology for quantum
cryptography have been reported [26] and the values of $\sigma _{x},\sigma
_{y},$ and $\sigma _{z}$ of a polarization qubit on a single photon can be
ascertained [27].Considered the experimental feasibility, this protocol can
be realized with today's technology. It is explained that when this paper
was completed, we see the protocol [28] presented by Deng $et$ $al$, which
also is a secure direct communication, using Einstein-Podolsky-Rosen pair
block.

I thank Yuan-chuan Zou for useful discussion. This work is supported by the
National Nature Science Foundation of China (Grant No. 10274094).

\section{References:}

[1] Nielsen M. A. and Chuang I. L. 2000 Quantum Computation and Quantum
Information (Cambridge University Press, Cambridge, UK)

[2] Peres A. 1998 Phys. Lett. A 128, 19; Duan L.-M. and Guo G.-C. 1998 Phys.
Rev. Lett., 80, 4999-5002

[3] Dieks D. 1982 Phys. Lett. A, 92, 271-272; Wootters W. K. and Zurek W. H.
1982 Nature, 299, 802-803; Barnum H., Caves C. M. , Fuchs C. A., Jozsa R.
and Schumacher B. 1996 Phys. Rev. Lett., 76, 2818-2821; Mor T. 1998 Phys.
Rev. Lett., 80, 3137-3140

[4] Bennett C. H. and Brassard G. 1984, in $proceedings$ $of$ $the$ $IEEE$ $%
International$ $Conference$ $on$ $Computers$, $Systems$ $and$ $%
\mathop{\rm Si}%
gnal$ $\Pr oces\sin g$, Bangalor, India, (IEEE, New York), p175-179.

[5] Ekert A. 1991 Phys. Rev. Lett. 67, 661

[6] Bruss D. 1998 Phys. Rev. Lett. 81, 3018

[7] Lo H. and Chau H. F. 1999 Science, 283, 2050

[8] Bostr$\stackrel{..}{o}$m K. and Felbinger T. 2002 Phys. Rev. Lett. 89,
187902

[9] Shor P. W. and Preskill J. 2000 Phys. Rev. Lett., 85, 441; Lo H. K.,
arXiv:quant-ph/0102138; Biham E., Boyer M., Boykin P. O., Mor T. and
Roychowdhury V 1999 arXiv:quant-ph/9912053.

[10] Schneier B. 1996 Applied Cryptography (Wiley, New York), 2nd ed.

[11] Holevo A. S. 1973 Statistical problems in quantum physics. In Gisiro
Maruyama and Jurii V. Prokhorov, editors, Proceedings of second Japan-USSR
Symposium on Probability Theory, pages 104-109, Springer-Verlag, Berlin

[12] Kraus K. 1983 States, Effects, and Operations (Spinger-Verlag, Berlin);
Barnum H., Nielsen M. A. and Schumacher B. 97 arXiv e-print
quantum-ph/9702049

[13] Shannon C. E. 1949 Bell Syst. Tech. J. 28, 656

[14] Cai Q.-Y. 2003 Phys. Rev. Lett., 91,109801

[15] Brassard G., Lutkenhaus N., Mor T. and Sanders B. C. 2000 Phys. Rev.
Lett., 85, 1330

[16] Bennett C. H., Bessette F., Brassard G., Salvail L. and Smolin J. 1992
J. Cryptology 5, 3

[17] de Martini F., di Guisippe G., and Marrocco M. 1996 Phys. Rev. Lett.,
76, 900

[18] Brouri R., Beveratos A., Poizat J-Ph. and Grangier P. 2000 Phys. Rev.
A, 62, 063817

[19] Lounis B. and Moerner W. E. 2000 Nature (London) 407, 491

[20] Treussart F., Clouqueur A., Grossman C. and Roch J.-F. 2001 Opt. Lett.
26, 1504

[21] Michler P., Imamouglu A., Mason M. D., Garson P. J., Strouse G. E. and
Buratto S. K. 2000 Nature (London) 406, 968

[22] Kim J., Benson O., Kan H. and Yamamoto Y. 1999 Nature (London) 397, 500

[23] Gisin N., Ribordy G., Tittel W. and Zbinden H. 2002 Rev. Mod. Phys. 74,
145

[24] Beveratos A., Brouri R., Gacoin T., Villing A., Poizat J.-P. and
Grangier P. 2002 Phys. Rev. Lett. 89, 187901

[25] Knill E., Laflamme R. and Milburn G. J. 2001 Nature (London) 409, 46 ;
Franson J. D., Donegan M. M., Fitch M. J., Jacobs B. C. and Pittman T. B.
2002 Phys. Rev . Lett., 89, 137901

[26] Ribordy G., Gautier J. D., Zbinden H. and Gisin N. 1998 Appl. Opt. 37,
2272-2277 ; Bourennane M., Gibson F., Karlsson A., Hening A., Jonsson P.,
Tsegaye T., Ljunggren D. and Sundberg E. 1999 Opt. Express 4, 383; Bethune
D. and Risk W. 2000 IEEE J, Quantum Electron. 36, 340 ; Hughes R., Morgan G.
and Peterson C. 2000 J. Mod. Opt. 47, 533; Ribordy G., Gautier J.-D., Gisin
N., Guinnard O. and Zbinden H. 2000 J. Mod. Opt. 47, 517

[27] Vaidman L., Aharonov Y. and Albert D. Z. 1987 Phys. Rev. Lett. 58, 1385
; Schulz O., Steinh$\stackrel{..}{u}$bl R., Weber M., Englert B. G.,
Kurtsiefer C. and Weinfurter H. 2003 Phys. Rev. Lett. 90, 177901

[28] Deng F.-G., Long G. L., and Liu X.-S. 2003 Phys. Rev. A 68 042317

\end{document}